\documentstyle[aps,pre]{revtex}
\begin{document}
\draft
\title{Steady-state structure of relativistic collisionless shocks}
\author{David Medvigy and Abraham Loeb}
\address{Harvard-Smithsonian Center for Astrophysics, 60 Garden Street,
Cambridge, MA 02138}
\date{\today}
\maketitle

\begin{abstract}
We explore analytically the structure of relativistic
shock and solitary wave solutions in collisionless plasmas.  In the wave
frame of reference, a cold plasma is flowing from one end and impacting on
a low velocity plasma. First we show that under
astrophysical conditions, a cold electron-positron plasma is unstable with
respect to a two-stream instability in the interface between these regions.
The instability heats the inflowing cold plasma rapidly, on a timescale
comparable to the inverse of its plasma frequency.
We then derive time-independent equations to describe the resulting hot
state of the pair plasma, and describe the conditions under which the
spatially uniform solution is the unique stable solution for the post shock
conditions.  We also examine plasmas composed of cold protons and hot
electrons, and show that the spatially uniform solution is the unique
stable solution there as well.  We state the shock jump conditions which
connect a cold, electron-proton plasma to a hot electron-proton plasma.
The generic feature evident in all of these models is that the plasma's
initial, directed kinetic energy gets almost completely converted into
heat.  The magnetic field plays the role of catalyst which can induce
the plasma instability, but our solutions indicate that the macroscopic
field only gets amplified by a factor of $\sim 3$ in the frame of the
shock.
\end{abstract}
\pacs{52.35.Tc, 52.35.Qz, 52.35.Sb, 52.60.+h}
\input epsf.sty

\section{Introduction}
Collisionless shock waves are common in astrophysical environments.  They
are a generic product of cosmic explosions, such as
supernovae\cite{helfand87,cargill88} 
(which result in non-relativistic shocks), and
gamma-ray bursts\cite{sari98,piran99} (which yield highly relativistic shocks).
Understanding their structure and dynamics is crucial for
modeling pulsar magnetospheres\cite{arons79}, active galactic
nuclei\cite{begel84}, structure formation in the intergalactic
medium\cite{loeb00}, and a local phenomenon such as the Earth's bow shock in
the solar wind\cite{tidmankrall}.  Despite the wide range of applications
related to collisionless shocks, their theoretical modeling 
is still at a primitive state, and so astrophysicists often resort to 
collisional fluid equations in modeling extra-solar systems with no
proper justification.

Because of the disparate physics used to describe 
collisional and collisionless plasmas, it is important to have an 
independent model of collisionless shocks.
In a collisional plasma, the inter-collision mean free 
path sets the scale for the
shock transition layer. Frequent 
collisions generate a Maxwellian distribution of post-shock particle 
energies, and all particle species eventually come to 
share a common temperature\cite{mckee80,draine93}.  
However, shocks also occur in plasmas for which binary collisions
are completely negligible\cite{langdon88,hoshino92,gallant94}.  
In this case, plasma instabilities act 
through macroscopic electromagnetic fields to 
bring about the shock wave, and possibly establish thermal 
equilibrium\cite{tidmankrall}.

Several aspects of collisionless shocks have been explored in the
literature. Steady, soliton solutions have been found\cite{alsop88}
for cold, magnetically dominated pair plasmas.  But when the plasma
kinetic energy is comparable to the Poynting energy,
particles are magnetically reflected and execute looping motions. 
Simulations have 
suggested\cite{langdon88,hoshino92,gallant92} that these 
looping motions are unstable, and
that a shock front forms in the region of magnetic reflection.  
Downstream of the shock, the plasma kinetic energy was seen to be mainly
converted into thermal energy. Since the simulations allowed 
particle motion in only two dimensions, it is unclear whether
or not the resulting momentum distribution is isotropic.
These simulations have been complemented by analytic 
studies\cite{weibel59,yoon87,medved99}  
of the two-stream instability, which is operative
whenever the fluid is composed of counter-moving streams.  
Since the aforementioned looping orbits present 
such a case, these analytical arguments suggest that the two-stream 
instability 
would dissipate the particles' ordered motion
on time scales comparable to the
inverse plasma frequency.  As the instability is saturated, the 
particles' momentum distribution isotropizes, 
and the magnetic field receives significant amplification.

In several astrophysical situations, and in particular the generation of
gamma-ray burst afterglows, it is important to understand what happens when
a relativistic shock wave impinges on a cold plasma carrying a 
small amount of energy in its
magnetic field, compared to the kinetic energy of the shock.  
In \S II we will investigate the
stability of a multi-stream configuration that typically results at
such an interface. We will derive a simple dispersion relation, valid
when the plasma frequency is much greater than the cyclotron frequency,
which indicates that this configuration is unstable.  This instability acts
on a time scale comparable to that observed in simulations, and provides a
promising driver for heating the gas and creating the necessary jump
conditions of collisionless, relativistic shocks.
The instability eventually saturates and converts the ordered bulk velocity
of the particles into a three-dimensional velocity distribution (thus,
playing a role similar to collisions in a normal fluid).  In \S III, we
will develop a stationary fluid model to describe the the post-shock
plasma.  After developing the appropriate equations, we will examine two
cases in particular.  First, we will investigate hot pair plasmas (\S IV),
and show that no stationary, continuous shock solutions exist.  
We will further derive the conditions under which oscillatory solutions
might exist, and show that they are similar to those obtained earlier for
cold plasmas\cite{alsop88}.  In \S V, we will investigate the case of an
electron-proton plasma.  Here, the two species have very different
dynamical length scales, and so we will construct a simple model in which
the electrons are hot, but the protons are cold.  As a first approximation,
we will average over the behaviour of the electrons, and calculate the
resulting fields and the motion of the protons.
Finally, we will consider a plasma in which both the electrons and the
protons are hot.  We will derived the jump conditions, and briefly discuss
the possibilities for spatial structure.  \S VI will summarize the main
conclusions of this work.

\section{The Instability}
\label{sect:two}
Previous studies have revealed that a generic feature of an initially
cold plasma carrying relatively little energy in its magnetic field is the
development of multiple
streams\cite{langdon88,hoshino92,gallant94,alsop88,medved99}.  The
condition that there be little energy in the magnetic field can be stated
in terms of the dimensionless ratio $\sigma\equiv B^2/8\pi\gamma^2 n m
c^2\ll 1$, where $B$ is the plasma's self-magnetic field, $\gamma$ is the
Lorentz factor corresponding to the mean fluid velocity, $n$ is the proper
number density, $m$ is the mass per particle, and $c$ is the speed of
light.  For simplicity, we suppose that the direction of the magnetic field
is initially perpendicular to the mean fluid velocity.  Then the definition
of $\sigma$ indicates that it is invariant with respect to Lorentz boosts
in the direction of the bulk flow.  This fact allows for a simple
calculation of $\sigma$ in the plasma's rest frame; for electrons in the
interstellar medium, $\sigma\approx 5\times 10^{-9}\left({B}/{\mu {\rm
G}}\right)^2 \left(n/{\rm cm^3}\right)^{-1}$.  This small value suggests
that multiple streams play an important role in 
the dynamics of shock and solitary
waves in the interstellar medium.

Simulations\cite{langdon88,hoshino92} have 
suggested that multi-stream situations are unstable,
and thus would require a time-dependent treatment.
We take an analytical approach to the question of stability 
by applying the collisionless Boltzmann equation, 
\begin{equation}
\label{eq:clbe}
\frac{\partial f}{\partial t}+{\bf v}\cdot \frac{\partial
f}{\partial {\bf x}}+e\left( {\bf E}+\frac{{\bf v}}{c}\times
{\bf B} \right) \cdot \frac{\partial f}{\partial {\bf p}}=0 ,
\end{equation}
to both species comprising a pair plasma.
Since $\sigma\ll 1$, the initial fields have 
negligible impact on length scales of the order of $c$/$\omega_p$,
where $\omega_p$ is the plasma frequency. Then we could analyze
the stability locally (over distances $\leq c/\omega_p$) with 
the fields completely neglected. 
Their primary role is to serve as
catalysts which induce multiple streams. With this, 
Eq.\ (\ref{eq:clbe}) becomes 
\begin{equation}
\label{eq:clbeb}
\frac{\partial f_0}{\partial t}+{\bf v}\cdot \frac{\partial
f_0}{\partial {\bf x}}=0.
\end{equation}
The initial two-stream distribution function 
satisfying Eq.\ (\ref{eq:clbeb}) is 
locally of the form
\begin{equation}
f_0(x_0,{\bf p})=\frac{n}{2}
\left[ \delta \left( p_x-p_{x0} \right) + 
\delta \left( p_x+p_{x0} \right) \right]
\delta \left( p_y\pm p_{y0} \right)
\delta \left( p_z \right) .
\label{eq:eqdf}
\end{equation}
The two species have values
of $p_{y0}$ that are equal in magnitude and opposite in sign because the 
magnetic field bends particles of opposite charge in opposite directions.
As a consequence, it is impossible to Lorentz boost to a frame where both
species have vanishing $p_{y0}$.
It is this nonzero value of $p_{y0}$ that distinguishes our
model from the classical two-stream instability.

We now calculate the effects of 
a perturbation in the fields and the distribution functions  
with the spacetime dependence $\exp i({\bf k\cdot x} - \omega t) $.  
Faraday's Law implies that the 
perturbed magnetic field is related to the perturbed 
electric field by ${\bf k}\times {\bf E} = \omega {\bf B}$.  
Using this in conjunction with Eq.\ (\ref{eq:clbe}) while
keeping only first-order terms in the perturbed quantities yields 
\begin{equation}
\left( -i\omega + i{\bf k\cdot v} \right) \delta f =
-e\frac{\partial f_0}{\partial {\bf p}}\cdot\left[ {\bf E}
+\omega ^{-1}{\bf v\times \left( k\times E \right)} \right] . 
\label{eq:perto}
\end{equation}
As suggested by\cite{medved99}, we specialize to the 
case in which {\bf E} is along the anisotropy axis, $\hat{x}$,  
and 
{\bf k} is along the $y$-direction.
Expanding the triple vector product, we find that 
\begin{equation}
\delta f = \frac{-ieE}{\omega - kv_y}\left[ \left(
1-\frac{k v_y}{\omega} \right) \frac{\partial f_0}{\partial p_x}
+ \frac{k v_x}{\omega}
\frac{\partial f_0}{\partial p_y}\right] .
\label{eq:pertb}
\end{equation}

For weak fields, the relation between {\bf E} and the displacement vector
{\bf D} is linear,
\( D_{\alpha}=\epsilon_{\alpha\beta} E_{\beta}. \) From this,
the definition of {\bf D} in terms of the polarization 
vector,
\( {\bf P}= \frac{1}{4\pi}\left( {\bf D}-{\bf E} \right) , \)
and the relation between the polarization and the current,
\( \partial {\bf P} / \partial t = {\bf J} \) , 
we obtain the following relation between the current and the
electric field:
\begin{equation} 
\sum_{species}\left[
e\int d^3{\bf p} v_{\alpha} \delta f \right] = 
\frac{-i\omega}{4\pi}
\left(\epsilon_{\alpha\beta}-\delta_{\alpha\beta}\right) E_{\beta} .
\label{eq:currefield}
\end{equation}
A scalar equation is obtained by contracting with
$E_{\alpha}$.  Since the initial fields are negligible
by assumption, the permittivity tensor may be decomposed 
into longitudinal and transverse parts. The
transverse piece is given by
\begin{equation}
\label{eq:transperm}
\left( \epsilon_{\alpha\beta} -\delta_{\alpha\beta} \right) E_{\beta}
= \left( \epsilon_t -1\right) E_{\alpha} .
\end{equation}
And so we obtain an equation for $\epsilon_{t}$:
\begin{equation}
\label{eq:epsit}
\frac{m\omega}{4\pi e^2}\left( \epsilon_t -1 \right) =
\int d^3 {\bf p} \frac{p_x}{\gamma\omega - k p_y / m}
\left[ \left( 1-\frac{k p_y}{\gamma m\omega} \right)
\frac{\partial}{\partial p_x} +\frac{k p_x}{\gamma m \omega}
\frac{\partial}{\partial p_y} \right] \sum_{species} f_0
\end{equation}

The delta functions make the integrations straightforward.  We 
integrate by parts to obtain
\begin{equation}
\label{eq:inteval}
\frac{m\omega}{4\pi e^2}\left( \epsilon_t -1 \right) =
\frac{-2n}{\gamma_0 \omega} \left( 1- v_{x0}^2 \right)
-\frac{2 nk^2 v_{x0}^2}{\gamma_0 \omega} \left( \frac{\omega^2
+k^2v_{y0}^2-3\omega^2 v_{y0}^2 + k^2 v_{y0}^4}{\left(
\omega^2-k^2 v_{y0}^2 \right) ^2} \right) ,
\end{equation}
where the velocities have been normalized by the speed of light. To 
obtain a dispersion relation for transverse, electromagnetic waves,
we set the permittivity tensor equal to $k^2/\omega^2$.
For simplicity we normalize the frequency in terms of the 
relativistic plasma frequency, $\omega_p$, and the wave number by 
$\omega_p /c$. Solving for $\omega$ leads to a cubic equation in $\omega^2$,
\begin{eqnarray}
0 & = & \omega^6 +\left( 2 v_{x0}^2 -2 - 2k^2 v_{y0}^2- k^2 \right) \omega^4
\nonumber \\ 
& & + \left( k^4 v_{y0}^4 +2k^4 v_{y0}^2 +4k^2v_{y0}^2 -2k^2 v_{x0}^2 
+2k^2 v_{x0}^2 v_{y0}^2 \right) \omega^2 
- k^6 v_{y0}^4 -2 k^4 v_{y0}^2 \left( v_{x0}^2 +v_{y0}^2 \right) .
\label{eq:komegrel}
\end{eqnarray}
This cubic depends on two parameters, viz., the
two components of  the velocity. 

\begin{figure}
\epsfxsize=4.0in\epsfbox{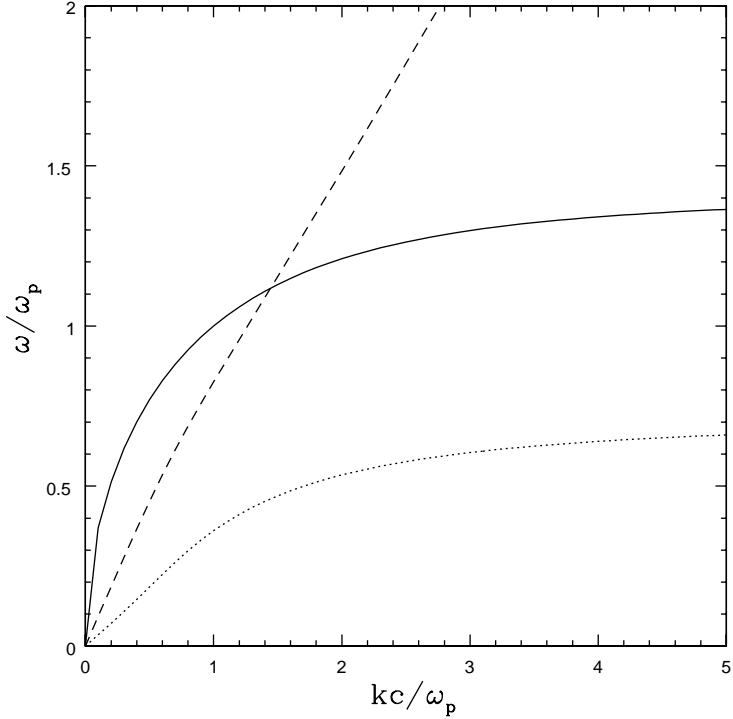}
\caption{The solid line depicts the growth rate of the standard two
stream instability as a function 
of wave number.  Each stream has a Lorentz factor of 100.  The 
dashed and dotted lines show the real and imaginary parts, respectively,
of the frequency for the modified two-stream instability discussed
in \S II. Here also the overall Lorentz factor 
is 100, but the $x$- and $y$-components of the velocity are 
equal in magnitude.   Only when the $y$-component is 
nonzero are both the real and imaginary parts of the 
frequency are nonzero.}
\label{fig:simplefreq}
\end{figure}

The exact solution to this equation is quite complicated and we
will not pause to write it in its entirety.  Instead, we
mention a few special cases.  First note that when $p_{x0}=0$,
the frequency is strictly real [See Eq.\ (\ref{eq:inteval})].  
Thus any instability must arise
from having overlapping streams of the {\it same} particle 
species.  The solution $\omega \left( k \right) $ is plotted 
for two cases in Fig.\ \ref{fig:simplefreq}.  As expected, when
$p_{y0}=0$, the frequency is either strictly real or strictly
imaginary.  This is the classic two-stream instability.  
As $p_{y0}$ becomes comparable to $p_{x0}$,  we see that the
plasma is still unstable; but in contrast to the previous
case, both real and imaginary parts of the frequency are nonzero.
The growth rate increases monotonically with $k$ and asymptotically 
approaches a number of the order of the plasma frequency, 
while the real part of the frequency grows approximately linearly
with $k$ when $ck\gg \omega_p$.  The instability is most powerful
for large $k$, or alternatively, over small distances.  This result
is self-consistent with our initial restriction of studying 
phenomena over distance scales shorter than $c/\omega_p$.  

\section{Basic Fluid Equations}
\label{se:eqn}

In the rest frame of a planar shock front, cold material is
flowing at a highly supersonic speed from $x=-\infty$.
The results  derived in the previous section
imply that a cold, low-$\sigma$ plasma inevitably
develops multiple streams and becomes unstable around
the shock front.  Simulations have shown
that the instability tends to heat the plasma
and eventually saturate\cite{langdon88,hoshino92}.
After saturation, we assume for simplicity that  the
collisionless plasma approaches a steady state.  In this section
we develop the time-independent equations necessary for the fluid
description of this state.

We consider the following boundary conditions at $x=-\infty$.
In the plasma's rest frame, we allow for a magnetic field in
$+\hat{z}$-direction.  A planar shock wave 
travels along the $\hat{x}$-direction with a
constant Lorentz factor, $\gamma_{sh}$.  We next boost to the rest frame of
the shock and assume that the plasma has reached a quiescent state, so that
all physical quantities are independent of time in this frame.  Moreover,
we take the shock front to be infinite in the $\hat{y}$- and
$\hat{z}$-directions; therefore, all physical quantities can only depend on
the $x$-coordinate, which represents the perpendicular distance to the
shock front. The Lorentz boost to the shock frame generates an electric
field in the $\hat{y}$-direction.  Other components of the fields and
velocities are assumed to be initially zero.

For the sake of simplicity, we work strictly with dimensionless quantities.  
Electromagnetic fields are normalized
by a reference value, $B_0$; spatial components of the four-velocity are
normalized by a reference value $\gamma_0 \beta_0 c$, while the time
component are normalized by $\gamma_0$.  Distances are normalized by
$\gamma_0 m_r c^2 / \left|e\right| B_0$ where $m_r$ is a reference mass to be
determined by the particle species of interest.  Finally, we define two
dimensionless variables \( \sigma= B_0^2/8\pi \gamma_0^2 m_r n_0 c^2\) and
\(T=P_0/n_0 m_r c^2 \). 
Note that these reference values are not,
in general, equal to the values given by the boundary conditions at
$x=-\infty$.
  
The equations of continuity can be obtained by integrating
Eq.\ (\ref{eq:clbe}) over momentum.  The resulting equations can be
immediately integrated yielding
\begin{equation}
\label{eq:continu} 
n u_x ={\rm constant} ,
\end{equation}
where $u_x$ is the $x$-component of the four-velocity.  Faraday's law shows
that the $y$- and $z$-components of the electric field are constant, and
can therefore be evaluated from the boundary conditions.  The condition,
${\bf \nabla}\cdot {\bf B}=0$, ensures that the $x$-component of the
magnetic field remains zero.  Poisson's equation determines how the
longitudinal electric field develops,
\begin{equation}
\frac{dE_x}{dx}=\frac{1}{2\sigma}\left( \frac{\gamma_i}{u_{xi}} -
\frac{\gamma_e}{u_{xe}} \right) .
\label{eq:poisson}
\end{equation}
The subscript $e$ 
denotes quantities pertaining to electrons, and the subscript
$i$ denotes quantities pertaining to the ions (either positrons or protons). 
The $x$-component of Ampere's Law is satisfied identically; the 
$y$ and $z$ components are 
\begin{equation}
\label{eq:ampy}
\frac{dB_y}{dx}=\frac{\beta_0}{2\sigma} 
\left( \frac{u_{zi}}{u_{xi}} -\frac{u_{ze}}{u_{xe}} \right) ,  
\end{equation}
\begin{equation}
\label{eq:ampere}
\frac{dB_z}{dx}=\frac{\beta_0}{2\sigma}
\left( \frac{u_{ye}}{u_{xe}} - \frac{u_{ye}}{u_{xe}} \right). 
\end{equation}

We assume that each species of particles has a stress-energy tensor of the
form $T^{\mu\nu}= diag(\rho , P , P ,P)$.  Though $P$ is not strictly the
pressure in a collisionless fluid, we will refer to it as such throughout
this paper for the sake of brevity.  To these we add the stress- energy
tensor of the electromagnetic field, and set the divergence of the sum
equal to zero to obtain the Euler equations [see Eqs.\
(\ref{eq:eulerx})-(\ref{eq:eulerz})].  Only three of these equations are
independent; therefore, we will only concern ourselves with the spatial
components.  To complete our set of equations, we need two equations of
state for the electron energy density and pressure, and two equations of
state for the correponding ionic quantities.  We assume that an
adiabatic description is valid, with adiabatic index $\Gamma$ between one
and two\cite{hoshino92,gallant94,kennel84}.  Then we
close our set of equations with
\begin{equation}
P=P_0 \left( \frac{n}{n_0} \right) ^{\Gamma}
\label{eq:adiaa}
\end{equation}
\begin{equation}
P=\left( \Gamma -1 \right) \left( \rho - mnc^2 \right) c^2
\label{eq:adiab}
\end{equation}
Eqs.\ (\ref{eq:continu}),(\ref{eq:adiaa}), and (\ref{eq:adiab}) 
allow us to eliminate $n$, $P$, and $\rho$, and 
so express the remaining equations completely in terms of the 
four-velocities and electromagnetic fields.
As described above, each species has three equations to 
describe its four-velocity, which are 
given by Eqs.\ (\ref{eq:eulerx})--(\ref{eq:eulerz}).
Because of our boundary conditions on $B_y$ and $u_z$, it is now clear
that they will remain zero. 

\begin{equation}
\left[ \frac{\Gamma T\left( 2-\Gamma \right) }
{\left( \Gamma - 1\right) u_x^{\Gamma -1 }}
+ \frac{m}{m_r} - \frac{\Gamma T}{\gamma_0^2\beta_0^2 u_x^{\Gamma +1}}
\right] \frac{du_x}{dx} = \frac{sgn(e)}{\beta_0^2}\left(\frac{\gamma E_x}
{u_x}
+B_z\beta_0\frac{u_y}{u_x} -B_y\beta_0\frac{u_z}{u_x}\right)
\label{eq:eulerx}
\end{equation}

\begin{equation}
\left[ \frac{\Gamma T}{\left( \Gamma - 1\right) u_x^{\Gamma -1 }}
+ \frac{m}{m_r} \right] \frac{du_y}{dx} = \Gamma T \frac{u_y}{u_x^{\Gamma}}
\frac{du_x}{dx} + \frac{sgn(e)}{\beta_0}\left(\frac{\gamma}{u_x}
-B_z\right)
\label{eq:eulery}
\end{equation}

\begin{equation}
\left[ \frac{\Gamma T}{\left( \Gamma - 1\right) u_x^{\Gamma -1 }}
+ \frac{m}{m_r} \right] \frac{du_z}{dx} = \Gamma T \frac{u_z}{u_x^{\Gamma}}
\frac{du_x}{dx} + \frac{sgn(e)}{\beta_0} B_y
\label{eq:eulerz}
\end{equation}

Our problem is to solve the remaining six, coupled, non-linear 
differential equations
for $B_z$ (or hereafter simply $B$), $E_x$, and the $x$ and $y$ 
components of the electron and ion four-velocities.  
This problem is simplified by the 
fact that three integrations can be immediately performed.
These represent the conservation of the $x$ and $y$ components of momentum
flux and the conservation of 
energy flux.  The equation corresponding to the conservation
of the $z$-component of momentum flux is satisfied identically.  
If $T^{\mu\nu}$ represents the combined energy-momentum
tensor of all species of particles and the fields, then these integrals
may be obtained most readily by realizing that 
$T^{\mu 1}=constant$.  The resulting equations are given below.

\begin{equation}
2\sigma \left( B-1\right) = \sum_{species}\left[
\frac{m}{m_r}\left(1-\gamma\right) + \frac{\Gamma T}{\Gamma-1}
\left( 1 - \gamma u_x^{1-\Gamma} \right) \right]
\label{eq:it10}
\end{equation}

\begin{equation}
\sigma \left( B^2 -1 +E_x^2\right) =  \sum_{species} \left[
\frac{m}{m_r}\beta_0^2\left( 1 -u_x \right)
+\frac{\Gamma T}{\Gamma -1}\beta_0^2\left( 1- u_x^{2-\Gamma} \right)
+\frac{T}{\gamma_0^2}\left( 1 - u_x^{-\Gamma} \right) \right]
\label{eq:it11}
\end{equation}

\begin{equation}
2\sigma E_x = \sum_{species}\left[
\frac{m}{m_r}\beta_0 u_y 
+\frac{\Gamma T}{\Gamma -1 }\beta_0 u_y u_x^{1-\Gamma} \right]
\label{eq:it12}
\end{equation}

\section{Pair Plasmas}
In this section we apply our model to pair plasmas.  We choose the reference
mass in the definition of $\sigma$ and $T$ to be the 
electron mass.  Because both species have the
same mass, the $x$-components of their four-momentum will be equal, and the
$y$-components will have equal magnitude and opposite sign.  From Eq.\
(\ref{eq:poisson}), we see that no longitudinal electric field will
develop.  Thus we need to determine spatial structure of the three
quantities $u_x$, $u_y$, and $B$, which are linked by Eqs.\ (\ref{eq:it10})
and (\ref{eq:it11}).  The simplest possible solution is one that is
constant in space. This solution can be used to assemble a simple picture
of a low-$\sigma$ collisionless shock. A plasma, with zero temperature at
$x=-\infty$, at some point encounters a shock wave generated by the
modified two-stream instability of Sec.\ \ref{sect:two}.  We expect the
shock width to be of order $c/\omega_p$ ( $\ll \left| e \right| B / \gamma
m c$), and so we approximate it as a discontinuity.  Then Eqs.\
(\ref{eq:it10})-(\ref{eq:it12}) may be used to obtain a constant solution
for the post-shock plasma\cite{kennel84}.  In situations where $\sigma\ll
1$, the jump conditions approach the classical hydrodynamic limit, given
by\cite{blandford76}, as might have been expected on the basis of our
equations of state.

Next we examine under what conditions the constant solution
is unique. Since the jump conditions of\cite{kennel84} were 
unique, the equations of a hot plasma 
cannot admit continuous shock wave solutions.  Solitary wave solutions
for a cold pair plasma were explored by\cite{alsop88}.  Here we  
extend their results to hot plasmas. 
From Eqs.\ (\ref{eq:it10}) and (\ref{eq:it11}), it is clear that 
all quantities must be bounded.  From 
Eq.\ (\ref{eq:it11}), we can consider 
$B^2$ to be a function of $u_x$. As $u_x$ goes to zero through
positive values, $B^2\rightarrow -\infty$.  As $u_x$ increases,
$B^2$ increases monotonically, eventually 
becomes positive, reaches a maximum, and then decreases 
monotonically.  Therefore the reality of $B$
implies the existence of a maximum and minimum value
of $u_x$, and also an upper limit on the magnitude of the 
magnetic field.  We do not consider the case where $u_x$ is less
than zero so as to avoid instabilities.

Eqs.\ (\ref{eq:eulerx}) 
and (\ref{eq:eulery}) show that the 
derivatives of the four-velocity diverge at a certain 
value of $u_x$.  This occurs because of the presence of the pressure
terms, and is absent in the zero-temperature case.  This
divergence might be remedied by more exact equations of state.  
The presence of this divergence, however, complicates 
the case under consideration.  This divergence occurs
at a value of $u_x$ such that $B$ is real; in fact, it occurs at
value of $u_x$ such that the right hand side of 
Eq.\ (\ref{eq:it11}) is maximized. So we must now not
only look for a bounded solution, but a bounded solution 
for which both turning points of $u_x$ are on the
same side of this singularity.  

On each side of the singularity, there is exactly one point where $B=0$.
Oscillatory solutions require at least two extrema of $u_x$, so it is
clear that 
we must find at least one extremum of $u_x$ where $u_y=0$.  But in 
fact we must find at least two. 
For suppose that there exists an oscillatory solution on the right of the 
singularity.  Then there must exist a minimum of $u_x$ such that 
$u_y=0$ in between the singularity and the point where $B=0$.  Examination
of the derivative of Eq.\ (\ref{eq:eulerx}) indicates that $B>0$
for a minimum of $u_x$ at such a point.  Since the 
only place where $B$ could go through zero is at a maximum of 
$u_x$, we see that if $B$ ever becomes less than zero there no 
longer exists the possibility of minimum, and so the `solution'
will inevitably become singular.  Applying similar reasoning 
to the left side of the singularity, we see that at any 
maximum of $u_x$, $B>0$.  We conclude as follows: any possible
oscillatory solutions must have $B>0$ everywhere.  As a 
corollary, there
must exist at least two turning points of $u_x$ (and also
$B$) where $u_y=0$.  

\begin{table}
\caption{Minimum values of $\sigma$
required for oscillating, steady-state solutions in a collisionless
pair plasma.  A value of $4/3$ is assumed for the adiabatic index.}
\label{tb:epmroot}
\begin{tabular}{llll} 
          & $\gamma_0=10$    &  $\gamma_0=100$  &  $\gamma_0=1000$  \\ \hline
$T$=0     & 4.6              &  50              & 5.1$\times 10^2$  \\ 
$T$=0.01  & 6.5              &  75              & 7.7$\times 10^2$  \\ 
$T$=1     & 39               & 2.9$\times 10^2$ & 1.6$\times 10^3$  \\ 
$T$=100   & 3.0$\times 10^3$ & 2.1$\times 10^4$ & 1.1$\times 10^5$  \\ 
$T$=1000  & 3.0$\times 10^4$ & 2.1$\times 10^5$ & 1.1$\times 10^6$  \\
\end{tabular}
\end{table}

In addition to these restrictions, we can show that there cannot
be any oscillatory solutions on the left of the the singularity.
At an extremum of $u_x$, both $u_y$ must vanish and the second
derivative of $u_x$ must have the sign appropriate to whether 
the extremum is a maximum or a minimum. 
This sign is completely determined by 
whether or not $B$ is greater or less than $\gamma / u_x$.  Since
we are only interested in $B>0$, we can equivalently compare
$B^2$ and $\gamma ^2/u_x^2$.  Note that the
right-hand side of this inequality is a monotonic decreasing
function of $u_x$, and the left hand side is given by Eq.\ (\ref{eq:it11}) 
and is montonic increasing in the region of interest.
Thus the two curves will intersect
either zero or one time. We
see that maxima could occur only to the left of this intersection,
while minima could occur only on its right.  Since there is no possibility
of having 
a maximum greater than a minimum, we must conclude that 
there exist no oscillatory solutions on the left of the 
singularity.

So for an oscillatory solution to exist, it must 
be on the right of the singularity, and there must be at least 
two points where $u_y=0$.  We can find any such points 
by equating Eq.\ (\ref{eq:it11}) and the square of 
Eq.\ (\ref{eq:it10}) with 
$u_y$ set equal to zero. For a given $u_x$, Eq.\ (\ref{eq:it11}) gives 
the actual value of $B$. Eq.\ (\ref{eq:it10}) gives only an 
upper limit since increasing $u_y^2$ can only decrease $B$.  Thus,
if there is an oscillatory solution in some range of $u_x$, the curve
defined by the square of  Eq.\ (\ref{eq:it10}) with $u_y=0$ must be 
greater than the curve defined by Eq.\ (\ref{eq:it11}) in the 
region between the extrema.   
The general solution of these equations, involving four parameters, 
$\sigma$, $\gamma_0$, $T$, and $\Gamma$, with $\Gamma$ not 
necessarily rational, cannot be expressed in terms of 
elementary functions.  Even if we take the
simpler cases of $\Gamma=4/3$ or $3/2$, corresponding to 
three or two dimensional relativistic velocity dispersion (at least in the
collisional case), we still must numerically solve a polynomial of 
high  degree in a three-dimensional parameter space. 
We therefore concentrate on the ultrarelativistic case.
Of particular interest is whether, for a given temperature and large Lorentz 
factor, there is a range of $\sigma$ which yields an
oscillating solution.  We examined this question numerically 
for $T=0,10^{-2},1,10^2$ and $10^4$. Our results, shown in 
Table~\ref{tb:epmroot}, indicate that oscillatory solutions
are indeed possible.  Similar to when $T=0$, only 
magnetically dominated plasmas exhibit 
these solutions.  The effect of a nonzero temperature is to
increase the minimum value of $\sigma$ for which a soliton 
solution emerges.  These results show that there are no
oscillatory solutions for $\sigma\ll 1$ in a relativistic 
pair plasma.  

\section{Electron-Proton Plasmas}
\subsection{Cold protons, hot electrons}
Next we investigate the steady-state properties of a plasma consisting of
cold protons and hot electrons.  In this section, we choose the 
reference mass in the definition of $\sigma$ and $T$ to be the 
proton mass.  As derived in Sec.\ \ref{se:eqn}, there
are only three independent quantities once Eqs.\
(\ref{eq:it10})-(\ref{eq:it12}) are accounted for.  As with pair plasmas, a
trivial steady-state solution exists in which all quantities remain
constant throughout space.  We now consider perturbations about this
solution.
 
Because the electron-to-proton mass ratio is very small, quantities
pertaining to the electrons vary considerably over a much shorter distance
than quantities pertaining to the protons.  In order to reduce the number
of dependent variables and make analytical progress, we begin by
investigating the spatial structure of the plasma on length scales of order
the proton Larmor radius.  We assume that the electrons undergo
oscillatory motion about the spatially uniform solution, and then
average the governing equations over several electron orbits.  Because the
motion of the electrons is bounded by assumption, we see that their
averaged velocity derivatives vanish.  Eqs.\ (\ref{eq:eulerx}) and
(\ref{eq:eulery}) for the protons go through essentially unchanged.  The
electrons' contribution to the right hand side of Eq.\ (\ref{eq:ampere})
vanishes, and their contribution to Eq.\ (\ref{eq:poisson}) is just a
constant number.  In Eqs.\ (\ref{eq:it10})--(\ref{eq:it12}) we 
average over powers of the electron bulk velocity.  So as not to
complicate the resulting equations, we take $<u_x^p>=<u_x>^p$.  Our
justification for doing this rests in the facts that $p\approx 1$ in almost
all of the terms, and in that we are looking for solutions that stay close
to the constant solution.  

With the above averaging, the problem is again reduced to solving for a
single dependent variable given two fixed parameters, $\sigma$ and $\gamma
_0$. Here $\sigma$ is defined in terms of the proton mass.  Our averaging
procedure evidently has the added advantage that there is no explicit
dependence on the electrons' equation of state.  To obtain a single
equation, we use Eq.\ (\ref{eq:it12}) to eliminate the electric field and
Eq.\ (\ref{eq:it11}) to eliminate $u_x$.  Eq.\ (\ref{eq:it10}) can then be
used to obtain a quadratic equation for $u_y^2$, with coefficients
depending on the magnetic field:
\begin{eqnarray}
0 & = & u_y^4+8\sigma\left( 2 \sigma -1 - \frac{\sigma}{\beta_0^2}
+ \frac{\sigma B^2}{\beta_0^2} \right) u_y^2
\nonumber \\
 & & + \frac{16\sigma^3}{\beta_o^2}\left[
\frac{\sigma}{\beta_0^2}-2-4\sigma+4\left( 1+2\sigma\right) B
-\left( 4\sigma +2+\frac{2\sigma}{\beta_0^2}\right) B^2
+\frac{\sigma}{\beta_0^2}B^4 \right] .
\label{eq:uyquad}
\end{eqnarray}
To eliminate $u_y$, we define a new independent variable
$d\tau = dx/u_x$.  Then, (the averaged) Eq.\ (\ref{eq:ampere}) illustrates 
that $u_y$ is simply proportional to the derivative of $B$.
To obtain extremal values of $B$, we set $u_y=0$ 
in Eq.\ (\ref{eq:uyquad}).
The result is a quartic polynomial 
which can be immediately deflated since $B=1$ is a double root.
The resulting quadratic can be solved to obtain the two additional 
roots
\begin{equation}
\label{eq:cubicb}
B=-1\pm\beta_0\sqrt{4+2/\sigma} .
\end{equation}

We now investigate what bounded solutions are possible. 
Eq.\ (\ref{eq:uyquad}) can easily be solved with the
quadratic formula for  
$\left( dB/d\tau \right) ^2$ yielding
\begin{equation}
\label{eq:uysqpm}
u_y^2 = -8\sigma^2+4\sigma+\frac{4\sigma}{\beta_0^2}
-\frac{4\sigma^2 B^2}{\beta_0^2}\pm
\frac{8\sigma^2}{\beta_0}\sqrt{2B^2-\left( 2+ \frac{1}{\sigma} \right) B
+\beta_0^2 \left( 1- \frac{1}{2\sigma} \right) ^2 + \frac{1}{\sigma} }
\end{equation}
 A sign ambiguity results;
however, once the sign of the radical is chosen, it will 
remain unchanged provided that $\beta_0 > 1/\sqrt{2}$.  
For lesser values of $\beta_0$, the radicand may go through zero; but 
since the zeros of the radicand are in 
general different from the extrema of $B$, this would
violate the condition that $u_y$ be real and there would
be no solution. Thus, once the sign of the radical it will 
remain fixed. 
Additional intuitive insight may be gained 
by considering $B$ to
be a generalized coordinate\cite{tidmankrall}.  Then 
the differential equation has
the form 
\( \frac{1}{2}\dot{B}^2 + \Phi \left( B \right) = 0 \) , i.e. 
$Kinetic\ Energy + Potential\ Energy = 0$. From this we can 
see that `energy' is conserved; thus, there will be no 
shock wave solutions.  

We now investigate the possibility of oscillatory 
solutions for $\sigma\ll 1$.  There are two possible extremal
values of $B$ (given by Eq.\ \ref{eq:cubicb}) which may be 
turning points of a physical solution.  First note that 
if $\sigma \ll 1$, then the root $B=1$ occurs when we choose the
minus sign in Eq.\ \ref{eq:uysqpm}.  To determine the sign which would
make the other two roots extrema, we substitute the values from
Eq.\ \ref{eq:cubicb} into Eq.\ \ref{eq:uysqpm}.  For $\beta_0\approx 1$ 
and $\sigma\ll 1$, the terms outside of the radical of Eq.\ \ref{eq:uysqpm}
are negative, indicating that the plus sign is the appropriate choice.
We can set these terms to zero to see what choice of parameters
results.  Solving for $\sigma$, we see that if $\beta_0 > 1/\sqrt{2}$, 
it is impossible to have $B$ given by Eq.\ \ref{eq:cubicb} be roots
of Eq.\ \ref{eq:uysqpm} with the minus sign.  Therefore, if a 
oscillating solution exists in the relativistic case, the 
extrema of $B$ are given by Eq.\ \ref{eq:cubicb} and we must choose 
the plus sign in Eq.\ \ref{eq:uysqpm}. 

However, the requirement that $u_x>0$ precludes the possibility of 
an oscillating solution.  For we have
\begin{equation}
\label{eq:uxprecl}
u_x=1-\frac{\sigma}{\beta_0^2}\left( B^2 -1 \right) - \frac{u_y^2}{4\sigma}.
\end{equation}
Provided $\beta_0>1/\sqrt{2}$, the point $B=1$ is in between the 
two extremal values of $B$.  Therefore, we evaluate Eq.\ \ref{eq:uxprecl}
for $B=1$.  This gives $u_x=-1+4\sigma$, which is certainly less 
than zero in many astrophysical situations where $\sigma \sim 10^{-9}$ (See
\S II).  Thus, there are no oscillating solutions, and the 
constant solution is unique.

\subsection{Hot Protons, Hot Electrons}
Finally, we consider the case where both the electrons and the protons are
hot, i.e. they both satisfy the equations of state~(\ref{eq:adiaa})
and~(\ref{eq:adiab}).  Again, we define $\sigma$ and $T$ in terms 
of the proton mass.  At $x=-\infty$ we take both species to be
cold, but at some later stage we suppose that they have undergone a
shock, in which both fluids were heated.  We can then find the uniform
post-shock solution as follows.  Eqs.\ (\ref{eq:poisson}) and
(\ref{eq:ampere}) show that both species have equal velocities.  Eq.\
(\ref{eq:eulerx}) shows that $u_y=0$ for both species, and therefore Eq.\
(\ref{eq:it12}) indicates that the longitudinal electric field vanishes.
Eqs.\ (\ref{eq:it10}) and (\ref{eq:it11}) may be used in conjuction with
Eq.\ (\ref{eq:eulery}) to determine $B$, $u_x$, and the total pressure.  It
is impossible from these considerations to determine how the total pressure
is split between the electrons and the protons.  Table~\ref{tb:eprjump}
shows the downsteam values of these quantities for several choices of
$\gamma_0$.  We have found the jump conditions to be independent of
$\sigma$ provided that $\sigma$ is less than approximately 0.1.

\begin{table}
\caption{The shock jump conditions for a low-$\sigma$, electron-proton plasma
for various shock Lorentz factors.
Upstream of the shock, the plasma is assumed to be cold.  But far 
downstream of the shock, the temperature has significantly increased. 
The jump conditions are essentially independent of $\sigma$, provided
$\sigma\ll 1$. The adiabatic index $\Gamma$ has been taken to be $4/3$.}
\label{tb:eprjump}
\begin{tabular}{llll} 
Lorentz Factor  & $B_{ds}/B_{us}$    &  $u_{x,ds}/u_{x,us}$  &  
$T_i + T_e = P_{0i}/n_0 m_p + P_{0e}/n_0 m_p$  \\ \hline
$10^2$ & 3.04 & 3.5$\times 10^{-3}$ & 3.54 \\ 
$10^3$ & 3.00 & 3.5$\times 10^{-4}$ & 16.7 \\ 
$10^4$ & 3.0  & 3.5$\times 10^{-5}$ & 77.4  \\ 
\end{tabular}
\end{table}

Analytical study of non-constant solutions of these 
differential equations is much more complicated than 
in the preceeding cases.  Here again, one must deal with
coupled differential equations.  One can try averaging over the
electrons' orbits, but solving for $u_{y,protons}$ is no longer 
straightforward, since it is no longer easy to eliminate
$u_{x,protons}$.  We have, however, numerically integrated these
equations with $\sigma\ll 1$ and $\gamma_0 \gg 1$, and have failed
to find satisfactory solutions.

\section{Conclusions}

We have examined the equations of a hot pair plasma, a plasma of 
hot electrons and cold protons, and finally a plasma in which 
both electrons and protons are hot.  In none of these cases 
were continuous shock wave solutions found. However, 
we have seen that a cold plasma with a small, embedded 
magnetic field is subject to a version of the 
two-stream instability, which is a likely
mechanism for the generation of collisionless shocks.
The standard two-stream instability can only
be saturated by nonlinear effects because of its aperiodic
nature, indicating that the magnetic field may be significantly
amplified\cite{medved99}.  But with this modified two-stream instability, both
the real and imaginary parts of the frequency are nonzero, and so 
kinetic
effects such as collisionless damping and resonance broadening
may play an important role in the eventual saturation. The
final level of magnetic field amplification is unclear.
Since the growth rate of the instability is much higher
than the cyclotron frequency, we expect this instability to be a dominant
influence in the formation of a collisionless shock.

We have also investigated the possible existence of soliton
solutions in a hot, pair plasma. As with  
cold plasmas, we have found that such solutions exist only
for $\sigma\gg 1$.  Thus, the only
time-independent solution is one that is independent of the
spatial coordinate. This
simple structure has been observed in numerical simulations
\cite{gallant92}, which do suggest spatially uniform solutions, with 
physical
quantities equal to the values predicted by the jump conditions.  

We have also considered a model consisting of hot electrons and cold
protons.  Since we have averaged over the electrons' orbits, this model is
essentially independent of the electrons' equations of state.  In this case
again we have found that the unique solution for a relativistic,
low-$\sigma$ plasma is the spatially uniform one.  Finally, we obtained the
jump conditions for the state of a hot plasma of electrons and protons that
was initially cold and had undergone a collisionless shock.  As in the
other cases, we have found that almost all of the plasma energy is
converted into pressure.  The magnetic field gets amplified by a factor of
$\sim 3$ in the shock frame, and the post-shock value of $\sigma$ is only
amplified by a factor of $\sim \gamma_{sh}$.  In many astrophysical
situation, this leaves the fraction of energy contained in the magnetic
field still far below the equipartition limit.  Numerical simulations of
relativistic collisionless shocks would be very useful in understanding
whether these jump conditions are appropriate, and whether or not the
adopted model of a quiescent, low-$\sigma$ plasma is adequate. In
particular, such simulations would also provide information on the relative
thermal Lorentz factors for the electrons and protons, which is
of fundamental
importance for models of the relativistic blast wave
in gamma-ray burst afterglows\cite{medved99}.

\begin{acknowledgements}
DM acknowledges support from an NSF graduate research fellowship.
This work was also supported by NASA grants NAG 5-7039 and NAG 5-7768, NSF
grant 9900077, and by a grant from the Israel-US BSF (for AL).
\end{acknowledgements}


\begin{references}
\bibitem{helfand87} D. J. Helfand and R. H. Becker, Astrophys. J., 
{\bf 314}, 203 (1987).
\bibitem{cargill88} P. J. Cargill and K. Papadopoulos, Astrophys. J., 
{\bf 329}, L29 (1988).
\bibitem{sari98} R. Sari, T. Piran, and R. Narayan, Astrophys. J., 
{\bf 497}, L17 (1998).
\bibitem{piran99} T. Piran, Physics Reports, {\bf 314}, 575 (1999).
\bibitem{arons79} J. Arons, Space Sci. Rev. {\bf 24}, 437 (1979).
\bibitem{begel84} M. Begelman, R. D. Blandford, and M. J. Rees,
Rev. Mod. Phys. {\bf 56}, 163 (1984). 
\bibitem{loeb00} A. Loeb and E. Waxman, Nature, {\bf in press} (2000).
\bibitem{tidmankrall} D. A. Tidman and N. A. Krall, 
{\it Shock Waves in Collisionless Plasmas} (John Wiley and
Sons, Inc., New York, 1971).
\bibitem{mckee80} C. F. McKee and D. J. Hollenbach, Ann. Rev. Astron.
Astrophs., {\bf 18}, 219 (1980).
\bibitem{draine93}B. T. Draine and C. F. McKee, Annu. 
Rev. Astron. Astroph., {\bf 31},373 (1993).
\bibitem{langdon88} A. B. Langdon, J. Arons, and C. E. Max, 
Phys. Rev. Lett., {\bf 61}, 779 (1988).
\bibitem{hoshino92} M. Hoshino, J. Arons, Y. A. Gallant, and A. B. Langdon, 
Astrophys. J.,{\bf 390}, 454 (1992).
\bibitem{gallant94} Y. A. Gallant and J. Arons, Astrophys. J.,
{\bf 435}, 230 (1994).
\bibitem{alsop88} D. Alsop and J. Arons, Phys. Fluids {\bf 31},
839 (1988).
\bibitem{gallant92} Y. A. Gallant, M. Hoshino, A. B. Langdon, J. Arons,
and C. E. Max, Astrophys. J., {\bf 391}, 73 (1992).
\bibitem{weibel59} E. S. Weibel, Phys. Rev. Lett., {\bf 2}, 83 
(1959).
\bibitem{yoon87} P. H. Yoon and R. C. Davidson, Phys. Rev. A, {\bf 35},
2718 (1987).
\bibitem{medved99}M. V. Medvedev and A. Loeb, Astrophys. J., 
{\bf 526}, 697 (1999).
\bibitem{kennel84} C. F. Kennel and F. V. Coroniti, 
Astrophys. J., {\bf 283}, 694 (1984).
\bibitem{blandford76} R. D. Blandford and C. F. McKee, Phys. Fluids,
{\bf 19}, 1130 (1976).

\end{references}
\end{document}